\documentclass[11pt]{article}
\usepackage{amsmath,amssymb,amsfonts} 
\usepackage{epsfig} \usepackage{latexsym,nicefrac,bbm}
\usepackage{xspace}
\usepackage{color,fancybox,graphicx,subfigure,fullpage}
\usepackage[top=1in, bottom=1in, left=1in, right=1in]{geometry}
\usepackage{tabularx}
\usepackage{hyperref} 
\usepackage{pdfsync}
\usepackage[boxruled]{algorithm2e}
\usepackage{multicol}
\usepackage{enumitem}
\usepackage{multirow}
\usepackage{url}
\usepackage{bm}
\usepackage{mathtools} 

\renewcommand{\epsilon}{\varepsilon}

\newcommand{\eps}{\varepsilon}

\newtheorem{theorem}{Theorem}[section]

\def\FullBox{\hbox{\vrule width 6pt height 6pt depth 0pt}}

\def\qed{\ifmmode\qquad\FullBox\else{\unskip\nobreak\hfil
\penalty50\hskip1em\null\nobreak\hfil\FullBox
\parfillskip=0pt\finalhyphendemerits=0\endgraf}\fi}

\usepackage[T1]{fontenc}
\usepackage{lmodern}

\title{An Introduction to Hamiltonian Monte Carlo Method for Sampling}
\author{Nisheeth K. Vishnoi}

\begin{document}

\maketitle
\begin{abstract}
The goal of this  article is to  introduce the Hamiltonian Monte Carlo method -- a Hamiltonian dynamics inspired algorithm for sampling from a Gibbs density $\pi(x) \propto e^{-f(x)}$.
We focus on the  ``idealized'' case, where one can compute continuous trajectories exactly.
We show that idealized HMC preserves $\pi$ and  we establish its convergence when $f$ is strongly convex and smooth.
\end{abstract}

\section{Introduction}
Many fundamental tasks in disciplines such as machine learning (ML), optimization, statistics, theoretical computer science, and molecular dynamics rely on the ability to  sample from a  probability distribution.
Sampling is used both as a tool to select training data and to generate samples and infer statistics from models such as highly multi-class classifiers, Bayesian networks,  Boltzmann machines, and GANs; see \cite{Gupta:2014,salakhutdinov2008bayesian,Koller:2009,Goodfellow-et-al-2016}.
 Sampling is also useful to provide robustness to optimization methods  used to train deep networks by allowing them to escape local minima/saddle points, and to prevent them from overfitting \cite{welling2011bayesian,dauphin2014identifying,salakhutdinov2008bayesian}.
Sampling methods are  often the key to solve integration, counting, and volume computation problems that are rampant in various applications at the intersection of sciences and ML \cite{dyer1991random}.     
In chemistry and molecular biology sampling is used to estimate reaction rates and simulate molecular dynamics which, in turn, is  used for discovering new materials and drugs \cite{carter1989constrained, rosso2002use, zheng2013rapid}. 
In finance, sampling is used to optimize expected return of portfolios \cite{detemple2003monte}. 
Sampling {algorithms} are also used to solve partial differential equations in applications such as geophysics \cite{cliffe2011multilevel}.

Mathematically, several of the above applications reduce to the following problem: given a function $f:\mathbb{R}^d \rightarrow \mathbb{R},$ generate independent samples from the distribution $\pi(x) \propto e^{-f(x)}$, 
 referred to as the  Gibbs or Boltzmann distribution.
 Markov chain Monte Carlo (MCMC) is one of the most influential frameworks to design algorithms to sample from Gibbs distributions  \cite{Goodfellow-et-al-2016}.
 Examples of MCMC algorithms include Random Walk Metropolis (RWM),  ball walk, Gibbs samplers,
Grid walks \cite{dyer1991random}, Langevin Monte Carlo \cite{roberts1998optimal}, and a physics-inspired meta-algorithm --  Hamiltonian Monte Carlo (HMC) \cite{duane1987hybrid} -- that subsumes several of the aforementioned methods as its special cases.
HMC was first discovered by physicists \cite{duane1987hybrid}, and was adopted with much success in ML  \cite{neal1992bayesian, warner1997bayesian, chen2014stochastic, betancourt2015hamiltonian}, and is currently the main algorithm used in the popular  software package \emph{Stan} \cite{carpenter2016stan}.

 HMC algorithms are inspired from a physics viewpoint and reduce the problem of sampling from the Gibbs distribution of $f$ to simulating the trajectory of a particle in $\mathbb{R}^d$ according to the laws of Hamiltonian dynamics where $f$ plays the role of a ``potential energy'' function.
We show here that the physical laws that govern HMC also endow it with the ability to take long steps, at least in the setting where one can simulate Hamiltonian dynamics exactly.
Moreover, we present  sophisticated discretizations of continuous trajectories of Hamiltonian dynamics that can lead to sampling algorithms for  Gibbs distribution for ``regular-enough'' distributions. 
%see \cite{mangoubi2018dimensionally}.
%

\section{Related Algorithms for Sampling}
We give a brief  overview of a number of different algorithms available for sampling from continuous distributions.  
In a typical MCMC algorithm, one sets up a Markov chain whose domain includes that of $f$ and whose  transition function  determines the motion of a particle in the target distribution's domain.

\paragraph{Traditional  algorithms: Ball Walk, Random Walk Metropolis.}
In the ball walk and related random walk metropolis (RWM) algorithms, each step of the Markov chain is computed by proposing a step $\tilde{X}_{i+1} = X_i + \eta v_i$,  where $v_i$ is uniformly distributed on the unit ball for the ball walk and standard multivariate Gaussian for RWM, and $\eta>0$ is the ``step size''.
 The proposal is then passed through a Metropolis filter, which accepts the proposal with probability $\min \left(\frac{\pi(\tilde{X}_{i+1})}{\pi(\tilde{X}_i)}, 1\right)$, ensuring that the target distribution $\pi$ is a stationary distribution of the Markov chain.

Some instances and applications of these algorithms include sampling from and computing the volume of a polytope \cite{dyer1991random, applegate1991sampling, lovasz1990mixing, lovasz1992randomized, lovasz1993random,lovasz2003hit} as well as sampling from a non-uniform distribution $\pi$ and the related problem of integrating functions \cite{roberts1997weak, lovasz2006fast}.  
One advantage of these algorithms is that they only require a membership oracle for the domain and a zeroth-order oracle for $f$.  
On the other hand, the running times of these algorithms are sub-optimal in situations where one does have access to information such as the gradient of $f$, since they are unable to make use of this additional higher-order information.
  To ensure a high acceptance probability, the step size $\eta$ in {the} RWM or ball walk algorithms must be chosen to be sufficiently small so that $\pi(X_i)$ does not change too much at any given step. 
   For instance, if $\pi$ is $N(0,I_d)$, then the optimal choice of step size is roughly  $\eta \approx \frac{1}{\sqrt{d}}$, implying that $\|\eta v_i\| \approx 1$ with high probability \cite{roberts1997weak}. 
   Therefore, since most of the probability of a standard spherical Gaussian lies in a ball of radius $\sqrt{d}$, one would expect all  of these algorithms to take roughly $(\sqrt{d})^2 = d$ steps to explore a sufficiently regular target distribution.

 \paragraph{Langevin algorithms.}
The Langevin algorithm makes use of the gradient of $f$ to improve on the provable running times of the RWM and {b}all walk algorithms.  
Each update is computed as $X_{i+1} = X_i - \eta \nabla f(X_i) + \sqrt{2 \eta} V_i$, where $V_1, V_2, \ldots \sim N(0,I_d)$. 
Recently, provable running time bounds have been shown for different versions of the Langevin algorithm for $m$-strongly log-concave distributions with $M$-Lipschitz gradient \cite{dalalyan2017theoretical, durmus2017nonasymptotic, cheng2017underdamped, durmus2017nonasymptotic}, including an  $\tilde{O}(\max(d \kappa, d^{\frac{1}{2}} \kappa^{1.5})  \log(\frac{1}{\epsilon}))$ {bound} for getting $\eps$ close to the target distribution $\pi$ in the ``total variation'' (TV) metric, where $\kappa= \frac{M}{m}$ \cite{dwivedi2018log}. 
This bound has been improved to roughly $d \kappa$ in \cite{LeeST21}.

  Running time bounds for Langevin in the non-convex setting have also recently been obtained in terms of isoperimetric constants for $\pi$ \cite{raginsky2017non}.  
   For sufficiently regular target distributions, the optimal step size of the Langevin algorithm can be shown to be $\eta \approx d^{-\frac{1}{4}}$ without Metropolis adjustment and $d^{-\frac{1}{6}}$ with Metropolis adjustment \cite{roberts1998optimal, neal2011mcmc}. Since the random ``momentum'' terms $V_1,V_2,\ldots$ are independent, the distance traveled by Langevin in a given number of steps is still roughly proportional to the square of its inverse numerical step size $\eta^{-1}$, meaning that the running time is at least $\eta^{-2} = d^{\frac{1}{2}}$ without Metropolis adjustment and $\eta^{-2} = d^{\frac{1}{3}}$ with the Metropolis adjustment.  
   The fact that the  ``momentum'' is discarded after each numerical step is therefore a barrier in further improving the running time of Langevin algorithms.

\section{Hamiltonian Dynamics}
We present an overview of the Hamiltonian dynamics and its properties that play a crucial role in the description and analysis of HMC.
Imagine a particle of unit mass in a potential well ${f}: \mathbb{R}^d \rightarrow \mathbb{R}$.
 If the particle has position $x\in \mathbb{R}^d$ and velocity $v \in \mathbb{R}^d$, its total energy is given by the function ${H}:\mathbb{R}^{2d}\rightarrow \mathbb{R}$ defined as $${H}(x,v) = {f}(x) + \frac{1}{2} \|v\|^2.$$
This function is called the Hamiltonian of the particle.
The Hamiltonian dynamics for this particle are: 

$$
\frac{{d}x}{{d}t} =  \frac{\partial H}{\partial v}   \ \ \mbox{and} \ \  \frac{{d}v}{{d}t} = -\frac{\partial H}{\partial x} .
$$
In the case where the Hamiltonian has a simple forms such as mentioned above, these equations reduce to:

$$\frac{{d}x}{{d}t} =  v  \ \ \mbox{and} \ \  \frac{{d}v}{{d}t} = -\nabla {f}(x).$$ 
For a starting configuration $(x,v)$, we denote the solutions to these equations by $(x_t(x,v),v_t(x,v))$ for $t \geq 0$.
Let the ``Hamiltonian flow'' $\varphi_t: (x, v) \mapsto  (x_t(x, v), v_t(x, v))$ be the position and velocity after time $t$ starting from $(x,v)$.

Interestingly,  these solutions satisfy a number of conservation properties. 

\begin{enumerate}

\item Hamiltonian dynamics conserves the Hamiltonian: ${H}(x(t), v(t)) = {H}(x(0),v(0))$ for all $t\geq 0$.
To see this note that, since $H$ does not depend on $t$ explicitly (or $\frac{\partial H}{\partial t}=0$), 
$$ \frac{dH}{dt} = \sum_{i \in [d]} \frac{dx_i}{dt} \frac{\partial H}{\partial x_i} + \frac{dv_i}{dt} \frac{\partial H}{\partial v_i} = 0.$$

\item Hamiltonian dynamics conserves the volume in the ``phase space.''
Formally, let $F =  \left(\frac{dx}{dt},\frac{dv}{dt}\right)$ be the vector field associated to the Hamiltonian in the phase space $\mathbb{R}^d \times \mathbb{R}^d.$
First note that the divergence of $F$ is zero:
\begin{eqnarray*}
\mathrm{div} F &=& \nabla \cdot F = \sum_{i \in [d]} \frac{\partial}{\partial x_i} \frac{d x_i}{dt} + \frac{\partial}{\partial v_i} \frac{d v_i}{dt} \\
& = &  \sum_{i \in [d]} \frac{\partial}{\partial x_i} \frac{\partial H}{\partial v_i} -\frac{\partial}{\partial v_i} \frac{\partial H}{\partial x_i} \\
&=& 0.
\end{eqnarray*}
Since divergence represents the volume density of the outward flux of a vector field from an infinitesimal volume around a given point, it being zero everywhere implies volume preservation.
Another way to see this is that divergence is the trace of the Jacobian of the map $F$, and the trace of the Jacobian is the derivative of the determinant of the Jacobian.
Hence, the trace being $0$ implies that the determinant of the Jacobian of $F$ does not change.

\item Hamiltonian dynamics of the form mentioned above are time reversible for $t \geq 0$: $$\varphi_{t}(x_t(x,v),-v_t(x,v))=(x,-v).$$

\end{enumerate}
\noindent
The underlying geometry and conservation laws have been generalized significantly in physics and have led to the area of symplectic geometry in mathematics \cite{da2008lectures}.

\section{Hamiltonian Monte Carlo}
To improve on the running time of the Langevin algorithms, one must find a way to take longer steps while still conserving the target distribution. 
Hamiltonian Monte Carlo (HMC) algorithms accomplish this by taking advantage of conservation properties of Hamiltonian dynamics.  
These conservation properties allow HMC to choose the momentum at the beginning of each step and simulate the trajectory of the particle for a long time. 
HMC is a large class of algorithms, and includes the Langevin algorithms and RWM as special cases.  

Each step of the HMC Markov chain $X_1,X_2,\ldots$ is determined by first sampling a new independent momentum $\xi \sim N(0,I_d)$, and then running Hamilton's equations for a fixed time $T$, that is $X_{i} = x_T(X_{i-1}, \xi)$.
   This is called the {\em idealized} HMC, since its trajectories are the continuous solutions to Hamilton's equations rather than a numerical approximation.

\medskip

\begin{algorithm}[H]
\caption{Idealized Hamiltonian Monte Carlo} \label{alg_HMC}
\KwIn{First-order oracle for  $f: \mathbb{R}^d \rightarrow \mathbb{R}$, 
an initial point $X_0 \in \mathbb{R}^d$, $T\in \mathbb{R}_{>0}$, $k \in \mathbb{N}$}

\For{$i = 1, \ldots, k$}{

Sample a momentum $\xi \sim N(0,I_d)$ 

Set $(X_i,V_i)=\varphi_T(X_{i-1},\xi)$

}
Output $X_k$

\end{algorithm}

\medskip
\noindent
Despite its simplicity, popularity, and the widespread belief that HMC is faster than its competitor algorithms in a wide range of high-dimensional sampling problems \cite{creutz1988global, HMC_optimal_tuning, neal2011mcmc, betancourt2014optimizing}, its theoretical properties are relatively  less understood compared to its older competitor MCMC algorithms, such as the Random Walk Metropolis \cite{mattingly2012diffusion} or Langevin \cite{durmus2016sampling, durmus2016sampling2, dalalyan2017theoretical} algorithms.
 Thus, for instance, it is more difficult to tune the parameters of HMC.
  Several papers have  have made progress in bridging  this gap, showing that HMC is geometrically ergodic for a large class of problems \cite{livingstone2016geometric, durmus2017convergence} and proving quantitative bounds for the convergence rate of the idealized HMC for special cases \cite{seiler2014positive, mangoubi2016rapid}.

These analysis benefit from the observation that there are various invariants that are preserved along the Hamiltonian trajectories, which in principle obviate the need for a Metropolis step, and raise the possibility of taking very long steps (large $T$).  
Using these invariance properties, we first show that the idealized HMC ``preserves'' the target density (Theorem \ref{thm_stationary_distribution}) and subsequently give a dimension-independent bound on $T$ when $f$ is strongly convex and smooth (Theorem \ref{thm_convergence}).
While not the focus of this article, in Section \ref{sec:Discrete}, we discuss different  numerical integrators (approximate algorithms to simulate Hamiltonian dynamics dynamics in the non-idealized setting) and bounds associated to them.

\section{Stationarity: HMC  Preserves the Target Density}

Recall that for $(x,v) \in \mathbb{R}^d \times \mathbb{R}^d$ and $T \geq 0$, $\varphi_T(x,v)$ denotes the position in the phase space of the particle moving according to the Hamiltonian dynamics with respect to a Hamiltonian $H(x,v)=f(x)+\frac{1}{2}\|v\|^2$. %
Let $\mu$ be the Lebesgue measure on $\mathbb{R}^d \times \mathbb{R}^d$ with respect to which all densities are defined.

\begin{theorem} \label{thm_stationary_distribution}
Let $f:\mathbb{R}^d \to \mathbb{R}$ be a differentiable function.
Let $T>0$ be the step size of the HMC.
Suppose $(X,V)$ is a sample from the density
$$ \pi(x,v) = \frac{e^{-f(x) - \frac{1}{2}\|v\|^2}}{\int  e^{-f(y) - \frac{1}{2}\|w\|^2} d\mu(y,w)}.$$
Then the density of $\varphi_T(X,V)$ is  $\pi$
for any $T\geq 0$.
Moreover the density of $\varphi_T(X,\xi)$, where $\xi \sim N(0,I_d)$ is also $\pi$.
Thus, the idealized HMC algorithms preserves $\pi$.
\end{theorem}

\noindent
The proof of this theorem heavily relies on the properties of Hamiltonian dynamics.
For $T \geq 0$, let $(\tilde{x},\tilde{v})=\varphi_T(x,y)$.
Then, time reversibility of Hamiltonian dynamics implies that
$$ (x,-v) = \varphi_{T}(\tilde{x},-\tilde{v}).$$
And, it follows from the preservation of Hamiltonian along trajectories that 
$$ H(x,-v) = H(\varphi_{T}(\tilde{x},-\tilde{v})).$$
Thus,  
\begin{eqnarray*}
 H(x,v) &=& f(x) + \frac{1}{2}\| v\|^2 \\ 
 & = & H(x,-v) \\
 & = & H(\varphi_{T}(\tilde{x},-\tilde{v})) \\
  & = & H(\tilde{x},-\tilde{v})\\
  & = & f(\tilde{x}) + \frac{1}{2}\|-\tilde{v}\|^2 \\
  & = & H(\tilde{x},\tilde{v}).
 \end{eqnarray*}
 Thus, $e^{-H(x,v)}$, the value of density associated to $(\tilde{x},\tilde{v})$, is the same as $e^{-H(\tilde{x},\tilde{v})}$.
 Let $\mu_*$ be the pushforward of $\mu$ under the map $\varphi_T$.
 The property that Hamiltonian dynamics preserves volume in phase space  implies that $\mu_*=\mu$ as the determinant of the Jacobian of the map $\varphi_T$ is $1$.
 Thus, the density $\pi$ remains invariant under $\varphi_T$:
 $$\tilde{\pi}(\tilde{x},\tilde{v}) = \frac{e^{-H(\tilde{x},\tilde{v})}}{\int e^{-H(y,w)} d\mu_*(y,w)} = \frac{e^{-H(\tilde{x},\tilde{v})}}{\int e^{-H(y,w)} d\mu(y,w)} = \pi(\tilde{x},\tilde{v}).$$
To see the second part, note that the marginal density of $v$ drawn from $\pi$ is the same as that of $N(0,I_d)$.
Thus, $\pi$ is an invariant density of the idealized HMC algorithm.

\section{Convergence: Running Time Bound for Strongly Convex and Smooth Potentials}
For densities $\pi_1$ and $\pi_2$ with the same base measure, the Wasserstein distance $W_2$ is defined to be the infimum, over all  joint distributions of the random variables $X$ and $Y$ with marginals $\pi_1$ and $\pi_2$, of the expectation of the squared Euclidean distance $\|X-Y\|^2$.

\begin{theorem} \label{thm_convergence}
Let $f:\mathbb{R}^d \to \mathbb{R}$ be a twice-differentiable function which satisfies
$ mI \preceq \nabla^2 f (x) \preceq M I$.
Let $\nu_k$ be the distribution of $X_k$ at  step $k \in \mathbb{Z}^\star$ from Algorithm \ref{alg_HMC}. Suppose that both $\nu_0$ and $\pi$ have mean and variance bounded by $O(1)$.
Then given any $\eps>0$, for $T= \Omega \left( \frac{\sqrt{m}}{M} \right)$ and $k = O\left((\frac{M}{m})^2 \log\frac{1}{\epsilon}\right)$, we have that $W_2(\nu_k,\pi) \leq \epsilon$.

\end{theorem}

\noindent
The bound in this theorem can be improved to $O\left(\frac{M}{m} \log\frac{1}{\epsilon}\right)$; see \cite{ChenV19}.
To prove Theorem \ref{thm_convergence} we  use the coupling method.
In addition to the idealized HMC Markov chain $X$ which is initialized at some arbitrary point $X_0$, to prove Theorem \ref{thm_convergence} we also consider another ``copy'' $Y = Y_0, Y_1, \ldots$ of the idealized HMC Markov chain defined in Algorithm \ref{alg_HMC}.
To initialize $Y$ we imagine that we sample a point $Y_0$ from the density $\pi$.
Since we have already shown that $\pi$ is a stationary density of the idealized HMC Markov chain (Theorem \ref{thm_stationary_distribution}), $Y_k$ will preserve the distribution $\pi$ at every step $k \in \mathbb{Z}^\ast$.

\begin{figure}
\begin{center}
\includegraphics[scale=0.4]{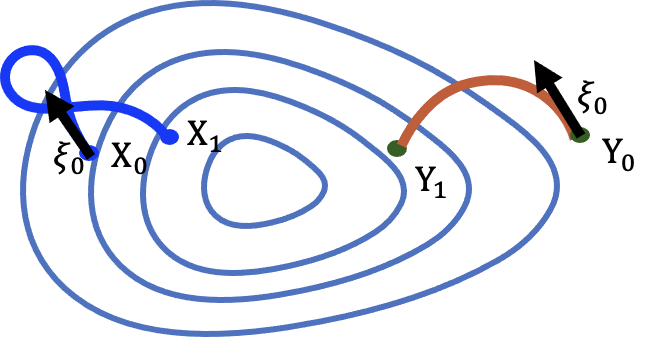}
\end{center}
\caption{ Coupling two copies $X$ (blue) and ${Y}$ (red) of idealized HMC by choosing the same momentum at every step. 
}\label{fig:coupling_Euler}
\end{figure}

To show that the density of $X$ converges to $\pi$ in the Wasserstein distance, we  design a coupling of the two Markov chains such that the distance between $X_k$ and $Y_k$ contracts at each step $k$.
If $\pi$ has mean and variance bounded by $O(1)$, and we initialize, e.g.,  $X_0 = 0$, then the we  have that $W_2(\nu_0, \pi) = O(1)$ as well since $Y_0 \sim \pi$.
Hence, $$W_2(\nu_0, \pi) \leq \mathbb{E}[\|Y_0-X_0\|^2]  = \mathbb{E}[\|Y_0\|^2] \leq \mathrm{Var}(Y_0) + \|\mathbb{E}[Y_0]\|^2 = O(1).$$
Thus, if we can find a coupling such that $\|X_k - Y_k\| \leq (1-\gamma) \|X_{k-1} - Y_{k-1}\|  $ for each $k$ and some $0<\gamma<1$, we would have that  $$W_2(\nu_k, \pi) \leq \mathbb{E}[\|X_k - Y_k\|^2] \leq (1-\gamma)^{2k} \mathbb{E}[\|X_0 - Y_0\|^2] = (1-\gamma)^{2k} \times O(1).$$
We  define a coupling of $X$ and $Y$ as follows:  At each step $i \geq 1$ of the Markov chain $X$ we sample an initial momentum $\xi_{i-1} \sim N(0,I_d)$, and set $(X_i,V_i) = \varphi_T(X_{i-1},\xi_{i-1})$.
And at each step $i \geq 1$ of the Markov chain $Y$ we sample momentum $\xi_{i-1}' \sim N(0,I_d)$, and set $(Y_i,V_i) = \varphi_T(Y_{i-1},\xi_{i-1}')$.
To couple the two Markov chains, we will give the same initial momentum  $\xi_{i-1}' \leftarrow \xi_{i-1}$ to the Markov chain $Y$; see Figure \ref{fig:coupling_Euler}.
This coupling preserves the marginal density of $Y$ since, in this coupling, $\xi_{i-1}'$ and $\xi_{i-1}$ both have marginal density $N(0,I_d)$.
Thus, we still have that the marginal density of $Y_i$ is $\pi$ at each step $i$.

\paragraph{Spherical harmonic oscillator.}

As a simple example, we  consider the spherical harmonic oscillator with potential function $f(x) \coloneqq \frac{1}{2} x^\top x$.
In this case the gradient is $\nabla f(x) = x$ at every point $x$.
Define $(x_t,v_t) \coloneqq \varphi_t(X_i,\xi_i)$ to be the position and momentum of the Hamiltonian flow which determines the update of the Markov chain $X$ at each step $i$, and $(y_t,u_t) \coloneqq \varphi_t(Y_i,\xi_i)$ to be the position and momentum of the Hamiltonian flow which determines the update of the Markov chain $Y$.
 Then we have $$\frac{{d} v_t}{{d} t} - \frac{{d} u_t}{{d} t} = -\nabla f(x_t)+ \nabla f(y_t)  = y_t - x_t.$$
 Thus, the difference between the force on the particle at $x_t$ and the particle at $y_t$ points in the direction of the particle at $y_t$.
 This means that, in the case of the spherical harmonic oscillator, after a sufficient amount of time $t$ the particle $x_t$ will reach the point $0$ and we will have $x_t- y_t = 0$.

We can solve for the trajectory of $x_t-y_t$ exactly.  We have
\begin{equation} \label{eq_spherical_harmonic_oscillator}
\frac{{d}^2(x_t-y_t)}{{d}t^2} = -(x_t - y_t).
\end{equation} Since the two Markov chains are coupled such that the particles $x$ and $y$ have the same initial momentum, $v_0 = u_0$,  the initial conditions for the ODE in Equation \eqref{eq_spherical_harmonic_oscillator} are $\frac{d (x_t - y_t)}{dt} = v_0 - u_0 = 0$.
Therefore, the solution to this ODE is
\begin{equation}
    x_t - y_t = \cos (t) \times (x_0- y_0).
\end{equation}
Hence, after time $t = \frac{\pi}{2}$ we have  $x_t - y_t = 0$.

\paragraph{General harmonic oscillator.}
 
 More generally, we can consider a harmonic oscillator $f(x) = \sum_{j=1}^d c_j x_j^2$, where $m\leq c_j \leq M$.
 In this case the gradient is $\nabla f(x) = 2C x$, where $C$ is the diagonal matrix with $j$-th diagonal entry $c_j$.

 In this case, we have $$\frac{{d} v_t}{{d} t} - \frac{{d} u_t}{{d} t} = -\nabla f(x_t)+ \nabla f(y_t)  = 2C(y_t - x_t).$$
 Thus, since $c_j>0$ for all $j$, the difference  $\frac{{d} v_t}{{d} t} - \frac{{d} u_t}{{d} t}$ between the force on the particle at force on the particle at $x_t$ and the particle at $y_t$ is has a component in the direction $y_t - x_t$. 
 This means that, for small values of $t$,  two particles will move towards each other at a rate of roughly $$\frac{1}{\|y_t - x_t\|}(y_t - x_t)^\top (2C)(y_t - x_t) t = \frac{1}{\|y_t - x_t\|} \|\sqrt{2C}(y_t - x_t)\|^2 t,$$ since the initial velocities $v_0$ and $u_0$ of the two particles are equal.
  However, unless $C$ is a multiple of the identity matrix, the difference between the forces also has a component orthogonal to $y_t - x_t$.
Thus, unlike in the case of the spherical harmonic oscillator, the distance between the two particles will not in general contract to zero for any value of $t$.

As in the case of the spherical harmonic oscillator, we can compute the distance between the two particles at any value of $t$ by solving for the trajectory $x_t - y_t$ exactly.
Here we have
\begin{equation}
\frac{{d}^2(x_t-y_t)}{{d}t^2} = -2C(x_t - y_t)
\end{equation}
with initial conditions $\frac{d (x_t - y_t)}{dt} = 0$.
Since $C$ is diagonal, the ODE is separable along the coordinate directions, and we have, for all $j \in [d]$
\begin{equation}
\frac{{d}^2(x_t[j]-y_t[j])}{{d}t^2} = -2 c_j(x_t[j] - y_t[j])
\end{equation}
with initial conditions $\frac{d (x_t[j] - y_t[j])}{dt} = 0$.
Therefore, $x_t[j] - y_t[j] = \cos(\sqrt{2 c_j} t) \times (x_0[j]- y_0[j])$.
Hence, since $c_j \leq M$, the distance $x_t[j] - y_t[j]$ will contract up to at least time $T = \frac{\pi}{2} \times \frac{1}{\sqrt{2 M}}$.
In particular, for any $j$ such that $c_j = M$ after time $T = \frac{\pi}{2} \times \frac{1}{\sqrt{2 M}}$ we have that $ x_T[j] - y_T[j] = 0$.
However, since the $c_j$ are values such that $m \leq c_j \leq M$, there is in general no single value of $T$ such that all $x_T[j] - y_T[j] = 0$.
However, we can show that for $T = \frac{\pi}{2} \times \frac{1}{\sqrt{2 M}}$,

\begin{align}
    x_T[j] - y_T[j] &= \cos\left(\sqrt{2 c_j} T\right) \times (x_0[j]- y_0[j])\\
    &\leq \cos\left(\sqrt{2 m} \times  \frac{\pi}{2} \times \frac{1}{\sqrt{2 M}}\right) \times (x_0[j]- y_0[j])\\
    &\leq \left(1-\frac{1}{8}\left(\sqrt{2 m} \times  \frac{\pi}{2} \times \frac{1}{\sqrt{2 M}}\right)^2\right) \times (x_0[j]- y_0[j])\\
    &\leq \left(1- \Omega\left(\frac{m}{M} \right)\right) \times (x_0[j]- y_0[j]) \qquad \forall j \in [d],
\end{align}
where the inequality holds because the fact that $f$ is $m$-strongly convex implies that $c_j \geq m$, $\cos$ is monotone decreasing on the interval $[0, \pi]$, and $\sqrt{2 m} \times  \frac{\pi}{2} \times \frac{1}{\sqrt{2 M}} \in [0, \pi]$ since $0<m \leq M$.
The second inequality holds because $\cos(s) \leq 1- \frac{1}{8}s^2$ for all $s \in [0,\pi]$.
Hence, we have that
\begin{align}
    \|x_T - y_T\| &\leq  \left(1- \Omega\left(\frac{m}{M} \right)\right) \times \|x_0- y_0\|.
\end{align}

\paragraph{General strongly convex and smooth $f$ (sketch).}

Recall that in the case of the spherical Harmonic oscillator, $f(x) = m x^\top x$, the difference between the forces on the two particles is
\begin{equation}
  -\nabla f(x_t)+ \nabla f(y_t)  = 2m(y_t - x_t)
\end{equation}
and thus, the difference between the force acting on $x_t$ and the force acting on $y_t$ is a vector which points exactly in the direction of the vector from $x_t$ to $y_t$.

In more general settings where $f$ is $m$-strongly convex and $M$-smooth, but not necessarily a quadratic/harmonic oscillator potential, strong convexity implies that the component of the vector $  -\nabla f(x_t)+ \nabla f(y_t)$ in the direction $(y_t - x_t)$ still points in the direction $(y_t - x_t)$ and still has magnitude at least $2m\|y_t - x_t\|$.
However, unless $m=M$, the difference in the forces, $-\nabla f(x_t)+ \nabla f(y_t)$, may also have a component orthogonal to the vector $y_t - x_t$.
Since $f$ is $M$-smooth, this orthogonal component has magnitude no larger than $2M \|y_t - x_t\|$.

Thus, since the two Markov chains are coupled such that the two particles have the same initial velocity, that is, $v_0 - u_0 = 0$, we have, in the worst case,
\begin{align}
    x_t - y_t &= x_0 - y_0 + (v_0 - u_0)\times t - m t^2(x_0 - y_0) + Mt^2 z \|x_0-y_0\| + \textrm{Higher-order terms}(t)
    \\
   &= x_0 - y_0 - m t^2(x_0 - y_0) + Mt^2 z \|x_0-y_0\| + \textrm{Higher-order terms}(t),
\end{align}
where $z$ is a unit vector orthogonal to $(x_0-y_0)$.
We would like to determine the value of $t$ which minimizes $\|x_t - y_t\|$, and the extent to which the distance $\|x_t - y_t\|$ contracts at this value of $t$.
In this proof sketch we will ignore the higher-order terms.
These higher-order terms can be bounded using comparison theorems for ordinary differential equations; see Section 4 of \cite{mangoubi_AOAP_HMC}. 

Ignoring the higher-order terms, we have (since $z$ is a unit vector orthogonal to $(x_0-y_0)$), 
\begin{align} \label{eq_second_order}
    \|x_t - y_t\|^2 &\leq (1- m t^2)^2\|x_0 - y_0\|^2 + (Mt^2)^2 \|x_0-y_0\|^2\\
    &=\left((1- mt^2)^2 + (Mt^2)^2\right)\|x_0-y_0\|^2\\
    &=\left(1- 2mt^2 + m^2 t^4 + M^2 t^4\right)\|x_0-y_0\|^2 \\
    &\leq \left(1- 2mt^2 + 2M^2t^4\right)\|x_0-y_0\|^2.
\end{align}
The RHS of \eqref{eq_second_order} is minimized at $t = \frac{\sqrt{m}}{\sqrt{2} M}$.
Thus, for $t = \frac{\sqrt{m}}{\sqrt{2}M}$ we have:
\begin{align}
    \|x_t - y_t\|^2 
    &\leq \left(1- 2m\left(\frac{\sqrt{m}}{\sqrt{2}M}\right)^2 + M^2\left(\frac{\sqrt{m}}{\sqrt{2}M}\right)^4 \right)\times  \|x_0 - y_0\|^2 \\
    &= \left(1- \frac{m^2}{2M^2}\right) \times  \|x_0 - y_0\|^2. 
\end{align}

\noindent
Thus, we have $\gamma = \frac{m^2}{2M^2}$ for $T = \Theta\left(\frac{\sqrt{m}}{M}\right)$.

\section{Discretizing HMC}\label{sec:Discrete}
To prove (non-asymptotic) running time bounds on HMC, we must approximate $x$ and $v$ in the idealized HMC with some numerical method.
One can use a numerical method such as the Euler \cite{griffiths2010numerical} or leapfrog integrators \cite{hairer2003geometric}.
The earliest theoretical analyses of HMC were the asymptotic ``optimal scaling'' results of \cite{kennedy1991acceptances}, for the special case when the target distribution is a multivariate Gaussian.  Specifically, they showed that the Metropolis-adjusted implementation of HMC with leapfrog integrator requires a numerical step size of $O^{*}(d^{-\frac{1}{4}})$ to maintain an $\Omega(1)$ Metropolis acceptance probability in the limit as the dimension $d \rightarrow \infty$.  They then showed that for this choice of numerical step size the number of numerical steps HMC requires to obtain samples from Gaussian targets with a small autocorrelation is $O^{*}(d^{\frac{1}{4}})$ in the large-$d$ limit.   \cite{pillai2012optimal}  extended their asymptotic analysis of the acceptance probability to more general classes of separable distributions.

The earliest non-asymptotic analysis of an HMC Markov chain was provided in \cite{seiler2014positive} for an idealized version of HMC based on continuous Hamiltonian dynamics, in the special case of Gaussian target distributions.
As mentioned earlier, \cite{mangoubi2017rapid} show that idealized HMC can sample from general $m$-strongly logconcave target distributions with $M$-Lipschitz gradient in $\tilde{O}(\kappa^2)$ steps, where $\kappa = \frac{M}{m}$ (see also \cite{Bou-Rabee2,bou2018coupling} for more  work on idealized HMC).
They also show that an unadjusted implementation of HMC with first-order discretization can sample with Wasserstein error $\epsilon>0$ in $\tilde{O}(d^{\frac{1}{2}} \kappa^{6.5} \epsilon^{-1})$ gradient evaluations.  
In addition, they show that a second-order discretization of HMC can sample from separable target distributions in $\tilde{O}(d^{\frac{1}{4}} \epsilon^{-1} f(m,M,B))$ gradient evaluations,  where $f$ is an unknown (non-polynomial) function of $m,M,B$, if the operator norms of the first four Fr{\'e}chet derivatives of the restriction of $U$ to the coordinate directions are bounded by $B$. 
\cite{lee2017convergence} use the conductance method to show that an idealized version of the Riemannian variant of HMC (RHMC) has mixing time with total variation (TV) error $\epsilon>0$ of roughly $\tilde{O}(\frac{1}{\psi^2 T^2} R \log(\frac{1}{\epsilon}))$, for any $0\leq T \leq d^{-\frac{1}{4}}$, where $R$ is a regularity parameter for $U$ and $\psi$ is an isoperimetric constant for $\pi$.
Metropolized variants of HMC have also been studied recently; see \cite{ChenDW020} and the references in there.

\paragraph{A second-order discretization.} Here we discuss the approach of \cite{mangoubi2018dimensionally}.  
They approximate a Hamiltonian trajectory with a second-order Euler integrator that iteratively computes second-order Taylor expansions $(\hat{x}_{\eta}, \hat{v}_{\eta})$ of Hamilton's equations, where 
$$ 
 \hat{x}_{\eta}(x,v) = x+v\eta -\frac{1}{2} \eta^2  \nabla {f}(x), \qquad \hat{v}_{\eta}(x,v) = v - \eta \nabla {f}(x)  -\frac{1}{2} \eta^2 \nabla^2f(x) v.
$$
\noindent
Here, $\eta>0$ is the parameter corresponding to the ``step size''.
If we approximate 
$$ \nabla^2f(x) v \approx \frac{\nabla f(\hat{x}_\eta) - \nabla f(x)}{\eta},$$
we obtain the following numerical integrator:
$$ 
 \hat{x}_{\eta}(x,v) = x+v\eta -\frac{1}{2} \eta^2  \nabla {f}(x), \qquad \hat{v}_{\eta}(x,v) = v   -\frac{1}{2} \eta \left(  \nabla f(x) - \nabla f(\hat{x}_\eta) \right) .
$$
This leads to the following algorithm.

\begin{algorithm}[H]
\caption{Unadjusted Hamiltonian Monte Carlo} \label{alg_UHMC}

\KwIn{First-order oracle for  $f: \mathbb{R}^d \rightarrow \mathbb{R}$, 
an initial point $X_0 \in \mathbb{R}^d$, $T\in \mathbb{R}_{>0}$, $k \in \mathbb{N}$, $\eta>0$}

\For{$i = 1, \ldots, k$}{

Sample $\xi \sim N(0,I_d)$. \\
Set $q_0=X_i$ and $p_0=\xi$ \\
\For{$j = 1, \ldots,  \frac{T}{\eta}$}{
Set $q_{j}= q_{j-1}+ \eta p_{j-1} - \frac{1}{2}\eta^2 \nabla f(q_{j-1})$ and $p_{j}=p_{j-1}-\frac{1}{2}\eta \left( \nabla f(q_{j-1}) -\nabla f(q_{j}) \right)$

}
Set $X_{i} = q_{\frac{T}{\eta}}(X_{i-1}, \xi)$

}
Output $X_k$

\end{algorithm}

\noindent
It can be shown that under a mild ``regularity'' condition, this unadjusted HMC requires at most (roughly) $O\left(d^\frac{1}{4}\eps^{-\frac{1}{2}}\right)$ gradient evaluations; see \cite{mangoubi2018dimensionally} for details.

\medskip
\noindent
{\bf Higher-order integrators.} More generally, one can replace the second-order Taylor expansion  with a $k$th-order Taylor expansion for any $k\geq 1$, to obtain a $k$th-order numerical integrator for Hamiltonian trajectories.
Unfortunately, the number of components in the Taylor expansion grows exponentially with $k$ because of the product rule, so it is difficult to compute the Taylor expansion exactly.
  However, it is possible to compute this expansion approximately using polynomial interpolation methods such as the ``Collocation Method'' 
  of \cite{lee2016geodesic}
that was used for the special case when $f$  is constant on a polytope \cite{lee2017convergence}.
While higher-order integrators can give theoretical bounds, they are generally unstable in practice as they are not symplectic.

Developing practical higher-order methods and identifying other interesting regularity conditions on the target density that lead to  fast algorithms remain interesting future directions.

\section*{Acknowledgments}
The author would like to thank Oren Mangoubi for his help  with the proof of Theorem \ref{thm_convergence} which initially appeared in \cite{mangoubi2017rapid}, and Yin-Tat Lee, Anay Mehrotra, and Andre Wibisono for useful comments. The author would also like to acknowledge the support of  NSF CCF-1908347.

\bibliographystyle{alpha}
\bibliography{hmc} 

\end{document}